\documentclass[12pt,thmsa]{article}
\usepackage{amssymb}

\usepackage{sw20lart}

\input{tcilatex}
\begin{document}

\author{Emilio Santos. Emeritus Professor of Theoretical Physics \and Departamento
de F\'{i}sica. Universidad de Cantabria. Santander. Spain \and \TEXTsymbol{<}%
santose@unican.es\TEXTsymbol{>}}
\title{Effects of the quantum vacuum at a cosmic scale and dark energy}
\date{November 2024}
\maketitle

\begin{abstract}
Einstein equation in a semiclassical approximation is applied to a spherical
region of the universe, with the stress-energy tensor consisting of the mass
density and pressure of the $\Lambda CDM$ cosmological model plus an
additional contribution due to the quantum vacuum. Expanding the equaation
in powers of Newton constant G, the vacuum contributes to second order. The
result is that at least a part of the acceleration in the expansion of the
universe may be due to the quantum vacuum fluctuations.
\end{abstract}

\section{ Dark energy, cosmological constant and vacuum fluctuations}

In this paper I will study the possible effects of the quantum vacuum at a
scale larger than the typical distances between galaxies. The work may
provide clues to ascertain whether vacuum fluctuations might be the origin
of dark energy.

The hypothesis of dark energy (DE) has been introduced in order to explain
the \textit{accelerating} expansion of the universe \cite{Riess}, \cite
{Perlmutter}, \cite{Turner}, \cite{Corredoira}. DE consists of a density and
pressure 
\begin{equation}
\rho _{DE}=-\left( 1+\varepsilon \right) p_{DE}\simeq (6.0\pm 0.2)\times
10^{-27}kg/m^{3},  \label{2.0}
\end{equation}
filling space homogeneously\cite{Bennett}, \cite{Kowalski}. The nature of
the dark energy is unknown but the empirical fact that $\left| \varepsilon
\right| <<1$ shows that its effect is fairly equivalent to a cosmological
constant \cite{Peebles}.

As is well known the cosmological constant (CC) was introduced by Einstein
in order to get a stationary (although not stable) model of the universe.
Later on the discovery of the expansion of the universe made the CC useless,
but it was a recurrent possibility for about 70 years although without too
much empirical support\cite{Weinberg}. The view changed in 1998 when it was
discovered that the expansion of the universe was accelerating \cite{Riess}, 
\cite{Perlmutter}, which might be seen as the effect of a CC, although the
less committed assumption has been made that the acceleration is caused by a
hypothetical ingredient named DE. In any case some of the proposals about
the nature of DE are similar to previous assumptions about the origin of a
possible CC.

An early proposal was that CC may correspond to the energy and pressure of
the quantum vacuum. If this was the case a plausible assumption seemed to be
that its value could be got via a combination of the universal constants $c,$
$
\rlap{\protect\rule[1.1ex]{.325em}{.1ex}}h%
,G$. There is a unique combination with dimensions of density, that is
Planck density with a value 
\begin{equation}
\rho _{Planck}=\frac{c^{4}}{G^{2}
\rlap{\protect\rule[1.1ex]{.325em}{.1ex}}h%
}\simeq 10^{97}kg/m^{3},  \label{Planckdensity}
\end{equation}
which is about 123 orders greater than either the known value of DE eq.$%
\left( \ref{2.0}\right) $ or any reasonable value for a CC$.$ This big
discrepancy has been named the ``cosmological constant problem''\cite
{Weinberg}, \cite{Padma}. Many proposals have been made in the past for the
origin of a CC \cite{Weinberg} (or DE, see e.g. \cite{Motta} and references
therein) which shall not be discussed here. One of them has been the quantum
vacuum origin as said above. If this is the case then some mechanism should
exist reducing eq.$\left( \ref{Planckdensity}\right) $ to eq.$\left( \ref
{2.0}\right) ,$ but the fine tuning required looks unplausible, even
conspiratory \cite{Weinberg}. However I point out that, although the mean
energy of the vacuum might cancel by some mechanism, the fluctuations cannot
cancel completely which suggets that the fluctuations could give rise to an
effective CC or DE.

Using dimensional arguments any theory aimed at explaining DE eq.$\left( \ref
{2.0}\right) $ would involve at least a new parameter, in addition to the
universal constants, c, 
\rlap{\protect\rule[1.1ex]{.325em}{.1ex}}h%
, G. If we choose the parameter to be a mass, $m$, then the value of the
dark energy could be written in the form (with $c=1$) 
\begin{equation}
\rho _{DE}\approx \frac{G^{n}m^{2n+4}}{
\rlap{\protect\rule[1.1ex]{.325em}{.1ex}}h%
^{n+3}},  \label{1}
\end{equation}
$n$ being a real number. The choice $n=-2$ would remove $m$ and lead to the
Planck density eq.$\left( \ref{Planckdensity}\right) ,$ but $n=1$ may give
the observed value eq.$\left( \ref{2.0}\right) $ for $\rho _{DE}$ provided
that $m$ is of order the pion mass. Indeed more than forty years ago
Zeldovich\cite{Zel} proposed a formula like eq.$\left( \ref{1}\right) $ with 
$n=1$ in order to get a plausible value for a cosmological constant.
Furthermore he interpreted the result in terms of the mass $m$ and its
associated ``Compton wavelength'' $\lambda $, as follows 
\begin{equation}
\Lambda \equiv \rho _{CC}\sim -\frac{Gm^{2}}{\lambda }\times \frac{1}{%
\lambda ^{3}},\lambda \equiv \frac{
\rlap{\protect\rule[1.1ex]{.325em}{.1ex}}h%
}{m}.  \label{Zeld}
\end{equation}
Thus eq.$\left( \ref{Zeld}\right) $ looks like the energy density
corresponding to the (Newtonian) gravitational energy of two particles of
mass $m$ placed at a distance $\lambda $, assuming that such an energy
appears in every volume $\lambda ^{3}$ (although in eq.$\left( \ref{Zeld}%
\right) $ the gravitational energy $\Lambda $ would be negative if both
masses are positive). Zeldovich interpretation was that the ``particles''
were actually vacuum fluctuations. Hence his hypothesis that a finite CC
might exist deriving from the fluctuations of the quantum vacuum. In recent
times some modifications of Zeldovich\'{}s proposal have been attempted as
an explanation for the dark energy, identifying the CC with DE eq.$\left( 
\ref{2.0}\right) $ \cite{Santos11}.

In the present paper I study again the possibility that vacuum fluctuations
do produce a gravitational effect similar to a DE. At a difference with
previous papers\cite{Santos11}, where heuristic arguments were used, here I
will use a more formal quantum approach to the vacuum fluctuations. Indeed
the quantum vacuum fluctuations are specific quantum features, therefore
classical equations like Friedman\'{}s (see next section) are not
appropriate in order to get the contents of the universe from the observable
value of the accelerated expansion. In summary a correct approach should
involve quantum field theory and general relativity.

We should deal with quantized general relativity, but no fully satisfactory
quantum gravity is available Then I will make some approximate or \textit{%
effective} approach to the gravity of a quantum system, that is the quantum
vacuum. In fact I will integrate a semi-classical Einstein equation of
general relativity approximated to second order in the Newton constant $G$,
as will be explained in section 3.2. An effective treatment of the quantum
vacuum will be studied in section 3.1.

\section{Revisiting the argument for dark energy}

Our quantum approach in section 3 will parallel the standard procedure to
get DE eq.$\left( \ref{2.0}\right) $ from the observed accelerated expansion
of the universe\cite{Riess}, \cite{Perlmutter}, \cite{Turner}. For this
reason I will revisit that method, which allows relating observable
properties of spacetime with the contents of the universe via general
relativity. Indeed astronomical observations are compatible with the
universe having a Friedmann-Lema\^{i}tre-Robertson-Walker (FLRW) metric with
flat spatial slices\cite{Sahni} of the form 
\begin{equation}
ds^{2}=-dt^{2}+a(t)^{2}\left[ dr^{2}+r^{2}(d\theta ^{2}+\sin ^{2}\theta
d\phi ^{2})\right] ,  \label{0}
\end{equation}
the parameter $a(t)$ taking into account the expansion of the universe.
Indeed at present time $t_{0}$, it is related to the measurable Hubble
constant, $H_{0}$, and deceleration parameter, $q_{0}$, via 
\begin{equation}
\left[ \frac{\stackrel{.}{a}}{a}\right] _{t_{0}}=H_{0},\text{ }\left[ \frac{%
\ddot{a}}{a}\right] _{t_{0}}=-H_{0}^{2}q_{0}.  \label{2}
\end{equation}

From the function $a(t)$ the contents of the universe may be obtained
solving the Friedman equation (which is a particular case of Einstein
equation appropriate for the FLRW metric). The result is that, asides from
the baryonic mass density, $\rho _{B}$, two hypothetical ingredients seem to
exists, namely an additional (dark) matter having mass density $\rho _{DM}$
with negligible pressure, and another component with positive energy
density, $\rho _{DE},$ but negative pressure $p_{DE}=-\rho _{DE},$ labeled
dark energy (DE). In fact the following relations are obtained, as proved
below 
\begin{eqnarray}
\left[ \frac{\stackrel{.}{a}}{a}\right] ^{2} &=&\frac{8\pi G}{3}\left( \rho
_{B}\left( t\right) +\rho _{DM}\left( t\right) +\rho _{DE}\right) , 
\nonumber \\
\frac{\ddot{a}}{a} &=&\frac{8\pi G}{3}\left( \frac{1}{2}\left[ \rho
_{B}\left( t\right) +\rho _{DM}\left( t\right) \right] -\rho _{DE}\right) ,\;
\label{2.00}
\end{eqnarray}
where small effects of radiation and matter pressure are neglected.

The baryonic density $\rho _{B}$ is well known from the measured abundances
of light chemical elements, which allows calculating $\rho _{DE}$ and $\rho
_{DM}$ from the empirical quantities $H_{0}$ and $q_{0}$ via comparison of
eqs.$\left( \ref{2.00}\right) $ with eq.$\left( \ref{2}\right) $. The result
may be summarized in the $\Lambda CDM$ model. In it baryonic matter density, 
$\rho _{B},$ represents about 4.6\% of the matter content while cold dark
matter ($CDM$) and dark energy (represented by the greek letter $\Lambda $)
contribute densities $\rho _{DM}\sim $ 24\%, and $\rho _{DE}\sim $ 71.3\%
respectively. The values obtained by this method agree with data from other
observations. For instance cold dark matter, in an amount compatible with $%
\rho _{DM},$ is needed in order to explain the observed (almost flat)
rotation curves in galaxies.

In this section I revisit the derived relation of the metric of spacetime,
at the cosmological scale, with the mass densities $\rho _{B}\left( t\right)
,\rho _{DM}\left( t\right) ,\rho _{DE}$ and pressure $p_{DE}=-\rho _{DE}$ of
the $\Lambda CDM$ model. The standard approach is to use the FLRW metric as
said above but for our purposes it is more convenient to deal with a metric
alternative to FLRW, eq.$\left( \ref{0}\right) $, using curvature
coordinates for spherical symmetry whose most general metric is as follows 
\begin{equation}
ds^{2}=g_{rr}\left( r^{\prime },t^{\prime }\right) dr^{\prime 2}+r^{\prime
2}(d\theta ^{2}+\sin ^{2}\theta d\phi ^{2})-g_{tt}\left( r^{\prime
},t^{\prime }\right) dt^{\prime 2}.  \label{30}
\end{equation}
This metric may be appropriate for a small enough region of the universe
around us, but large in comparison with typical distances between galaxies%
\cite{Rich}.

The relation between the metrics eqs.$\left( \ref{0}\right) $ and $\left( 
\ref{30}\right) $ is as follows. We perform a change of variables in eq.$%
\left( \ref{0}\right) ,$ that is

\begin{equation}
r=a^{-1}r^{\prime },t=t^{\prime }-\frac{\dot{a}}{2a}r^{2},\dot{a}\equiv 
\frac{da}{dt},  \label{31}
\end{equation}
so chosen that, after some algebra, eq.$\left( \ref{0}\right) $ becomes eq.$%
\left( \ref{30}\right) $ where 
\begin{eqnarray}
g_{rr}\left( r^{\prime },t^{\prime }\right) &=&1+\left( \frac{\dot{a}}{a}%
\right) ^{2}r^{\prime 2}+O\left( r^{\prime 3}\right) ,  \nonumber \\
g_{tt}\left( r^{\prime },t^{\prime }\right) &=&1+\frac{\ddot{a}}{a}r^{\prime
2}+O\left( r^{\prime 3}\right) ,\ddot{a}\equiv \frac{d^{2}a}{dt^{2}}.
\label{32}
\end{eqnarray}
The calculation may be performed to order $r^{\prime 2},$ consistent with
the metric eq.$\left( \ref{30}\right) $ being appropriate for a small region
around us.

Now I shall solve (the classical) Einstein equation for the metric eq.$%
\left( \ref{30}\right) $ with a stress-energy tensor given by the $\Lambda
CDM$ model as described above, that is the mass (or energy) density$\ \rho
_{mat}$ of matter may be taken as the sum of two \textit{homogeneous}
contributions, that is a $\rho _{mat}=\rho _{B}\left( t\right) +\rho
_{DM}\left( t\right) ,$ meaning baryonic and dark matter respectively, with
negligible pressure, plus a dark energy with homogeneus density $\rho _{DE}$
and negative pressure $p_{DE}=-\rho _{DE}.$

The metric eq.$\left( \ref{30}\right) $ requires spherical symmetry, that is
both mass density and pressure should depend only on the radial coordinate $%
r $ and time $t$. Obviously this is not the case for the actual universe
where matter is mainly localized in galaxies. In practice an approximation
consists of averaging the mass density over the whole region. I point out
that a similar approximation is usually made when the FRW metric eq.$\left( 
\ref{2.0}\right) $ is used\cite{Rich}. The result of solving Einstein
equation for the metric eq.$\left( \ref{30}\right) $ is 
\begin{eqnarray}
g_{rr} &=&1+\frac{8\pi G}{3}\left[ \rho _{B}+\rho _{DM}+\rho _{DE}\right]
r^{2}+O\left( r^{3}\right) ,  \nonumber \\
g_{tt} &=&1+\frac{8\pi G}{3}\left[ \frac{1}{2}\left[ \rho _{B}+\rho
_{DM}\right] -\rho _{DE}\right] r^{2}+O\left( r^{3}\right) ,\;
\label{rofluct}
\end{eqnarray}
as is well known\cite{Rich}. Comparison of eqs.$\left( \ref{rofluct }\right) 
$ with eqs.$\left( \ref{32}\right) $ leads to eq.$\left( \ref{2.00}\right) .$

Now I proceed to the proof of eq.$\left( \ref{rofluct }\right) .$ I will
neglect terms of order $O\left( r^{3}\right) $ and ignore the (slow) change
of the metric coefficients with time, a change derived from the slow time
dependence of the matter density $\rho _{mat}$. With $g_{rr}=g_{tt}=1$ for $%
r=0$ we get the following elements for a metric like eq.$\left( \ref{30}%
\right) $\cite{Synge} 
\begin{eqnarray}
g_{rr}\left( r\right) &=&\left( 1-\frac{2Gm\left( r\right) }{r}\right) ^{-1},%
\text{ }m\left( r\right) =m_{mat}\left( r\right) +m_{DE}\left( r\right) 
\nonumber \\
m_{mat}\left( r\right) &=&\int_{\left| \mathbf{z}\right| \mathbf{<}r}\rho
_{mat}d^{3}z,m_{DE}\left( r\right) \equiv \int_{\left| \mathbf{z}\right| 
\mathbf{<}r}\rho _{DE}d^{3}z,  \nonumber \\
g_{tt}(r) &=&\exp \gamma ,\gamma =2G\int_{\left| \mathbf{x}\right| <r}\frac{%
m\left( x\right) +4\pi x^{3}p_{DE}\left( x\right) }{x^{2}-2Gxm\left(
x\right) }dx.  \label{4.0}
\end{eqnarray}
As typically $Gm\left( r\right) <<r$ an approximation is appropriate
consisting of expanding eqs.$\left( \ref{4.0}\right) $ in powers of the
Newton constant $G,$ retaining terms up to order $O\left( G^{2}\right) $.
For eqs.$\left( \ref{4.0}\right) $ this approximation agrees with order $%
O\left( r^{2}\right) $ in the radial parameter $r,$ as may be easily
checked. Thus I may write

\begin{equation}
g_{rr}=1+\frac{2Gm\left( r\right) }{r}+\frac{4G^{2}m\left( r\right) ^{2}}{%
r^{2}}+O\left( G^{3}\right) .  \label{gr}
\end{equation}
\begin{eqnarray}
g_{tt} &=&1+2G\int_{0}^{r}\left( \frac{m\left( x\right) }{x^{2}}+4\pi
xp\left( x\right) \right) dx+2G^{2}\left[ \int_{0}^{r}\left( \frac{m\left(
x\right) }{x^{2}}+4\pi xp\left( x\right) \right) dx\right] ^{2}  \nonumber \\
&&+4G^{2}\int_{0}^{r}m\left( x\right) \left( \frac{m\left( x\right) }{x^{3}}%
+4\pi p\left( x\right) \right) dx+O\left( G^{3}\right) .\smallskip
\label{gt}
\end{eqnarray}

Terms of order $O\left( G^{2}\right) $ will be relevant when the quantum
vacuum is taken into account, as in the quantum approach of the next section
but here we may neglect those terms$.$ As said above I take the contents of
the universe into account as in the $\Lambda CDM$ model, that is a
(homogeneous) mass density given by $\rho _{B}+\rho _{DM}+\rho _{DE}$ and
pressure $p_{DE}=-\rho _{DE}.$ Then eqs.$\left( \ref{gr}\right) $ and $%
\left( \ref{gt}\right) $ give eqs.$\left( \ref{rofluct}\right) .$

\section{A quantum treatment}

The aim of this article is to improve the standard derivation of the
contents of the universe from astronomical observations, essentially the
Hubble constant and the acceleration parameter as in eqs.$\left( \ref{0}%
\right) $ to $\left( \ref{2.00}\right) .$ This involves a quantized, rather
than classical, Einstein equation and the inclusion of the quantum vacuum
fluctuations. I shall start dealing with the approach to the quantum vacuum
and a semi-classical approximate Einstein equation of general relativistic.

\subsection{The quantum vacuum}

Vacuum fluctuations are straightforward consequences of quantum field theory
(QFT), but their treatment is far from trivial. In a naive approach the
energy of the vacuum is divergent, for instance in the quantum
electromagnetic field the vacuum energy in a finite volume is $%
E_{vac}=\sum_{j}
\rlap{\protect\rule[1.1ex]{.325em}{.1ex}}h%
\omega _{j}$ which goes to $\infty $ when we take all possible normal modes $%
j$ into account.. There are procedures to avoid the divergence, from the
simplistic ``normal ordering'' rule to the sophisticated renormalization
methods in quantum electrodynamics, the latter extremely successfull as is
well known. In the study of the influence of vacuum fluctuations at a cosmic
scale at least two alternatives arise. We may assume that the vacuum energy
density is just given by eq.$\left( \ref{2.0}\right) $, with the associated
pressure, thus explaining the nature of the dark energy. Other possibility
is that the vacuum energy density is strictly zero, the dark energy having a
different origin, maybe unrelated to the quantum vacuum. An intermediate
possibility is that just a part of DE is due to the quantum vacuum. In any
case it is natural within quantum theory to assume that the vacuum energy is
an observable that I should represent by a quantum operator $\hat{\rho}%
_{vac}\left( \mathbf{r},t\right) .$ Its vacuum expectation may be either
finite (positive), e.g. given by eq.$\left( \ref{2.0}\right) ,$ or just
zero. In this paper I attempt to study the latter posibility. Indeed if it
is finite it should be either all or a part of DE, but no further study of
that possibility will be made here.

Thus I shall assume the following

\begin{equation}
\hat{p}_{vac}\left( \mathbf{r},t\right) =-\hat{\rho}_{vac}\left( \mathbf{r}%
,t\right) ,\left\langle vac\left| \hat{\rho}_{vac}\left( \mathbf{r},t\right)
\right| vac\right\rangle =0,\left\langle vac\left| \hat{\rho}_{vac}\left( 
\mathbf{r},t\right) ^{2}\right| vac\right\rangle >0,\smallskip \smallskip
\smallskip  \label{1.0}
\end{equation}
where I have included the vacuum pressure operator, the former equality
deriving from the requirement of Lorentz invariance which is plausible.
Indeed we are considering a spacetime very close to Minkowski. The
inequality in eq.$\left( \ref{1.0}\right) $ means that the vacuum energy
density fluctuates. I point out that eqs.$\left( \ref{1.0}\right) $ might be
derived in principle from quantum field theory, including the vacuum
fluctuations of all fields (belonging to the standard model of high energy
physics) and their interactions.

For our purposes eqs.$\left( \ref{1.0}\right) $ are not sufficient in order
to characterize the quantum vacuum. The quantities relevant for our work are
the two-point correlations of the density and pressure of the vacuum. In an
approximate flat (Minkowski) space the vacuum should be invariant under
translations and rotations, whence the vacuum expectation of the product of
two vacuum density operators (at equal times) should be a universal
function, $C$, of the distance $\left| \mathbf{r}_{1}\mathbf{-r}_{2}\right| $%
\textbf{, }that is\textbf{\ } 
\begin{equation}
\frac{1}{2}\langle vac\left| \hat{\rho}_{vac}\left( \mathbf{r}_{1}\right) 
\text{ }\hat{\rho}_{vac}\left( \mathbf{r}_{2}\right) +\text{ }\hat{\rho}%
_{vac}\left( \mathbf{r}_{2}\right) \hat{\rho}_{vac}\left( \mathbf{r}%
_{1}\right) \right| vac\rangle =C\left( \left| \mathbf{r}_{1}\mathbf{-r}%
_{2}\right| \right) .  \label{3.4}
\end{equation}
The function C may be named self-correlation of the vacuum energy density.

In this article I will assume that the integral of $C(x)$ extended over the
whole space, is nil that is 
\begin{equation}
\int_{\left| \mathbf{r}_{2}\right| \in \left( -\infty ,\infty \right)
}C\left( \left| \mathbf{r}_{1}\mathbf{-r}_{2}\right| \right)
d^{3}r_{2}=\int_{\left| \mathbf{r}\right| \in \left( -\infty ,\infty \right)
}C\left( \left| \mathbf{r}\right| \right) d^{3}r=0.  \label{3.5}
\end{equation}
I believe that this assumption is plausible once we assume $\left\langle
vac\left| \hat{\rho}_{vac}\left( \mathbf{r},t\right) \right|
vac\right\rangle =0$ as in eq.$\left( \ref{1.0}\right) .$ The opposite
assumption might be worth to study, but will not be made here. The
implication 
\[
\left\langle vac\left| \hat{\rho}_{vac}\left( \mathbf{r},t\right) \right|
vac\right\rangle =0\Rightarrow \int \hat{\rho}_{vac}\left( \mathbf{r}%
,t\right) d^{3}r=0, 
\]
is similar to the ergodic property in (classical) stochastic processes.

Eq.$\left( \ref{3.5}\right) $ may be generalized to the self-correlation of
the pressure and the cross-correlation of density and pressure to be
introduced in section 4.2.

\subsection{Semiclassical approximation to a quantum Einstein equation}

We must solve the Einstein equation of GR for a stress-energy tensor that
should be defined for an essentially quantum system involving the quantum
vacuum. However there is no fully satisfactory quantum gravity theory that
unifies GR and quantum mechanics and we must rely in approximations. The
left side of Einstein equation of GR is known only in classical form, that
is the Einstein tensor $G_{\mu \nu }$. But the right side should be an
stress-energy tensor defined in quantum form, that is an operator $\hat{T}%
_{\mu \nu }$ in the Hilbert space. An equation with a classical tensor on
one side and a quantum operator tensor on the other is inconsistent and the
standard solution to the problem is the semiclassical approximation, that is 
\begin{equation}
G_{\mu \nu }=-8\pi G\left\langle \psi \left| \hat{T}_{\mu \nu }\right| \psi
\right\rangle ,  \label{3}
\end{equation}
where $\mid \psi \rangle $ is the quantum state of the system. Then both
sides of the (Einstein) equation are c-numbers. The Einstein tensor is a
function of the metric elements and their first and second derivatives with
respect to the coordinates. Therefore eq.$\left( \ref{3}\right) $ is a
partial differential equation involving the metric elements which in our
case consists of just two non-trivial ones, that is g$_{rr}$ and g$_{tt},$
see eq.$\left( \ref{30}\right) .$

The stress-energy tensor operator $\hat{T}_{\mu \nu }$ may be written,
similar to its classical counterpart of section 2, in terms of energy
density and pressure operators. In the classical case the contributions to
the mass (or energy) density, and pressure are given by the $\Lambda CDM$
cosmological model. That is mass densities $\rho _{mat}=\rho _{B}+\rho _{DM}$
and $\rho _{DE},$ and the pressure $p_{DE}=-\rho _{DE}.$ In eq.$\left( \ref
{3}\right) $ these quantities should be considered expectation values, in
the state $\mid \psi \rangle ,$ of appropriate quantum observables, that is 
\[
\left\langle \psi \left| \hat{\rho}_{mat}\right| \psi \right\rangle =\rho
_{mat},\left\langle \psi \left| \hat{\rho}_{DE}\right| \psi \right\rangle
=\rho _{DE},\left\langle \psi \left| \hat{p}_{DE}\right| \psi \right\rangle
=p_{DE}, 
\]
so that eq.$\left( \ref{3}\right) $ becomes just the classical Einstein
equation solved in section 2, if we ignore the possible quantum vacuum
contribution.

The semiclassical apparoximation eq.$\left( \ref{3}\right) $ presents the
problem that any information about fluctuations is lost. Indeed fluctuations
of a quantum observable, say $\hat{M}$, appear only in the higher moments $%
\left\langle \psi \left| \hat{M}^{n}\right| \psi \right\rangle $, $n\geq 2.$
However in our case all operators (of either mass density or pressure)
appear linearly in the stress-energy tensor $\hat{T}_{\mu \nu }$ whence
fluctuations are neglected in eq.$\left( \ref{3}\right) $. Thus in order to
study the effect of vacuum fluctuations I propose a different semiclassical
approximation, that is using the \textit{expectation value of an integrated,
rather than differential, Einstein equation }defined in eq.$\left( \ref{3}%
\right) $\textit{. }That semiclassical approximation may be represented as
follows 
\begin{equation}
g_{\mu \nu }=\left\langle \psi \left| \hat{g}_{\mu \nu }\left( \left\{ \hat{T%
}_{\lambda \sigma }\right\} \right) \right| \psi \right\rangle ,  \label{E}
\end{equation}
where $\left\{ g_{\mu \nu }\right\} $ are the elements of the metric and $%
\left\{ \hat{g}_{\mu \nu }\right\} $ the corresponding operators in a
hypothetical quantum Einstein equation.

Eq.$\left( \ref{E}\right) $ would be useless except if we were able to get
the function $\hat{g}_{\mu \nu }\left( \left\{ \hat{T}_{\lambda \sigma
}\right\} \right) $ of the operators $\hat{g}_{\mu \nu }$ in terms of the
elements of the operator tensor $\hat{T}_{\lambda \sigma }.$ This will be
possible just in some extremely simple situations. One of them is our case
of the metric eq.$\left( \ref{30}\right) $ with just two non-trivial metric
elements and where we may neglect the time dependence of the stress-energy
tensor. Then the two functions $\hat{g}_{rr}\left( \left\{ \hat{T}_{\lambda
\sigma }\right\} \right) $ and $\hat{g}_{tt}\left( \left\{ \hat{T}_{\lambda
\sigma }\right\} \right) $ will be similar to eqs.$\left( \ref{4.0}\right) $
with two modifications. Firstly we shall substitute the quantum operators $%
\hat{\rho}_{mat},$ $\hat{\rho}_{DE}$ and $\hat{p}_{DE}$ for the (classical)
expectations $\rho _{mat},\rho _{DE}$ and $p_{DE},$ respectively. Secondly
we shall include the vacuum operators $\hat{\rho}_{vac}\left( \mathbf{r}%
,t\right) $ and $\hat{p}_{vac}\left( \mathbf{r},t\right) $.

The result is the following eq.$\left( \ref{E}\right) $, consisting of just
two metric elements written in terms of the quantum operators for the energy
density and the pressure, 
\begin{eqnarray}
g_{rr}\left( r\right) &=&\left\langle \psi \left| 1+\frac{2G\hat{m}\left(
r\right) }{r}+\frac{4G^{2}\hat{m}\left( r\right) ^{2}}{r^{2}}\right| \psi
\right\rangle +O\left( G^{3}\right) ,  \nonumber \\
g_{tt}\left( r\right) &=&\left\langle \psi \left| 1+2G\int_{0}^{r}x^{-2}\hat{%
m}\left( x\right) dx+G^{2}\sum_{n=1}^{5}\hat{c}_{n}\right| \psi
\right\rangle +O\left( G^{3}\right) ,  \label{8.0}
\end{eqnarray}
where we define 
\begin{eqnarray}
\text{ }\hat{m}\left( r\right) &=&\hat{m}_{mat}\left( r\right) +\hat{m}%
_{DE}\left( r\right) +\hat{m}_{vac}\left( r\right) ,\hat{m}_{vac}\left(
r\right) \equiv \int_{\left| \mathbf{z}\right| \mathbf{\leq }r}\hat{\rho}%
_{vac}d^{3}z  \nonumber \\
\hat{m}_{mat}\left( r\right) &\equiv &\int_{\left| \mathbf{z}\right| \mathbf{%
\leq }r}\hat{\rho}_{mat}d^{3}z=\int_{\left| \mathbf{z}\right| \mathbf{>}r}%
\hat{\rho}_{mat}d^{3}z,\hat{m}_{DE}\left( r\right) \equiv \int_{\left| 
\mathbf{z}\right| \mathbf{\leq }r}\hat{\rho}_{DE}d^{3}z.  \label{7.0}
\end{eqnarray}
I will label $\left\{ \hat{c}_{n}\right\} $ the operators corresponding to
the terms of the sum eq.$\left( \ref{gt}\right) $ which now are promoted to
be operators, that is 
\begin{eqnarray}
\hat{c}_{1} &=&4\int_{0}^{r}x^{-3}\hat{m}\left( x\right) ^{2}dx,  \nonumber
\\
\hat{c}_{2} &=&\int_{0}^{r}x^{-2}dx\int_{0}^{r}y^{-2}dy\left[ \hat{m}\left(
x\right) \hat{m}\left( y\right) +\hat{m}\left( y\right) \hat{m}\left(
x\right) \right] ,  \nonumber \\
\hat{c}_{3} &=&32\pi ^{2}\int_{0}^{r}xdx\int_{0}^{r}ydy\left[ \hat{p}\left(
x\right) \hat{p}\left( y\right) +\hat{p}\left( y\right) \hat{p}\left(
x\right) \right] ,  \nonumber \\
\hat{c}_{4} &=&8\pi \int_{0}^{r}\left[ \hat{m}\left( x\right) \hat{p}\left(
x\right) +\hat{p}\left( x\right) \hat{m}\left( x\right) \right] dx, 
\nonumber \\
\hat{c}_{5} &=&8\pi \int_{0}^{r}x^{-2}dx\int_{0}^{r}ydy\left[ \hat{m}\left(
x\right) \hat{p}\left( y\right) +\hat{p}\left( y\right) \hat{m}\left(
x\right) \right] .  \label{13}
\end{eqnarray}

In the passage from eqs.$\left( \ref{gr}\right) $ and $\left( \ref{gt}%
\right) ,$ consisting of numerical quantities (c-numbers), to eqs.$\left( 
\ref{8.0}\right) ,\left( \ref{7.0}\right) $ and $\left( \ref{13}\right) $
involving operators, the problem appears of the ordering of operators that
not commute in general. In our case there are at most two operators in the
products eqs.$\left( \ref{13}\right) $ and they appear in symmetrical order,
which is most plausible.

Actually the solution of eq.$\left( \ref{3}\right) $ presents a difficulty
similar to the classical eqs.$\left( \ref{30}\right) $ and $\left( \ref{4.0}%
\right) .$ That is the solution eqs.$\left( \ref{8.0}\right) $ to $\left( 
\ref{13}\right) $ are valid only if the energy-momentum tensor operator $%
\hat{T}_{\mu \nu }$ depends on the coordinate $r$ but not on the angular
coordinates, $\theta ,\phi $. I will solve the problem as in the classical
case, eqs.$\left( \ref{gr}\right) $ and $\left( \ref{gt}\right) ,$ that is
averaging the energy density over large enough regions. However in the
quantum domain the solution is more involved. In fact in the classical
domain the dynamical variables are directly observables while in the quantum
domain the dynamical variables are represented by operators (usually labeled
``observables'') and the actually observable quantities are the expectation
values of the ``observables'' in the appropriate state $\mid \psi \rangle $.
Thus I will assume that the semiclassical eqs.$\left( \ref{8.0}\right) $ to $%
\left( \ref{13}\right) $, are valid when we deal with regions having
dimensions much larger than typical distances between galaxies, as is the
case in our work.

The metric element $g_{rr}$ consists of the following two terms 
\[
g_{rr}=g_{rr}^{\text{model}}+g_{rr}^{vac}, 
\]
where the superindex ``model'' stands for $\Lambda CDM$ model. The former
term will be calculated to order $O\left( G\right) $ because the
contribution of order $O\left( G^{2}\right) $ is negligible (see comment
below eq.$\left( \ref{4.0}\right) )$. The latter (vacuum term) should be got
to order $O\left( G^{2}\right) $ because the term of order $O\left( G\right) 
$ is nil, see eq.$\left( \ref{1.0}\right) $. Then to $O\left( G\right) $ we
get 
\begin{eqnarray}
g_{rr} &=&1+\frac{2G}{r}\int_{\left| \mathbf{z}\right| \mathbf{<}r}\left(
\left\langle \psi \left| \hat{\rho}_{\text{model}}\left( \mathbf{z}\right)
\right| \psi \right\rangle +\left\langle \psi \left| \hat{\rho}_{vac}\left( 
\mathbf{z}\right) \right| \psi \right\rangle \right) d^{3}z  \label{9.0} \\
&=&1+\frac{2G}{r}\int_{\left| \mathbf{z}\right| \mathbf{<}r}\left\langle
\psi \left| \hat{\rho}_{\text{model}}\left( \mathbf{z}\right) \right| \psi
\right\rangle d^{3}z,  \nonumber
\end{eqnarray}
which will reproduce the standard result, i.e. the first eq.$\left( \ref
{rofluct}\right) ,$ because that term involves 
\[
\left\langle \psi \left| \hat{\rho}_{\text{model}}\left( \mathbf{z}\right)
\right| \psi \right\rangle =\rho _{\text{model}}=\rho _{B}+\rho _{DM}+\rho
_{DE}. 
\]
Similarly the expectation of $\hat{g}_{tt}$ to order $O\left( G\right) $
will reproduce the second eq.$\left( \ref{rofluct}\right) .$

\subsection{Contribution of the quantum vacuum}

Taking eq.$\left( \ref{8.0}\right) $ into account, the term of order $%
O\left( G^{2}\right) $ of the $g_{rr}$ metric element is 
\begin{eqnarray}
g_{rr}^{vac} &=&\frac{4G^{2}}{r^{2}}\left\langle \psi \left| \hat{m}%
_{vac}\left( r\right) ^{2}\right| \psi \right\rangle =\frac{4G^{2}}{r^{2}}%
\left\langle \psi \left| \left[ \int_{\left| \mathbf{z}\right| \mathbf{<}r}%
\hat{\rho}_{vac}\left( \mathbf{z}\right) d^{3}z\right] ^{2}\right| \psi
\right\rangle  \nonumber \\
&=&\frac{2G^{2}}{r^{2}}\int_{\left| \mathbf{z}\right| \mathbf{<}%
r}d^{3}z\int_{\left| \mathbf{v}\right| \mathbf{<}r}d^{3}v\langle \psi \left| 
\hat{\rho}_{vac}\left( \mathbf{v}\right) \text{ }\hat{\rho}_{vac}\left( 
\mathbf{z}\right) +\text{ }\hat{\rho}_{vac}\left( \mathbf{z}\right) \hat{\rho%
}_{vac}\left( \mathbf{v}\right) \right| \psi \rangle .  \label{9}
\end{eqnarray}
Generalizing eq.$\left( \ref{3.5}\right) $ I assume that the two-point
correlation function, $C$, depends only on the distance $\left| \mathbf{v-z}%
\right| $\textbf{, }that is\textbf{\ } 
\begin{equation}
\frac{1}{2}\langle \psi \left| \hat{\rho}_{vac}\left( \mathbf{v}\right) 
\text{ }\hat{\rho}_{vac}\left( \mathbf{z}\right) +\text{ }\hat{\rho}%
_{vac}\left( \mathbf{z}\right) \hat{\rho}_{vac}\left( \mathbf{v}\right)
\right| \psi \rangle =C\left( \left| \mathbf{v-z}\right| \right) ,
\label{0.9}
\end{equation}
which implies in particular that we may neglect the possible perturbations
of the vacuum correlations due to the presence of matter. The function $%
C\left( \left| \mathbf{v-z}\right| \right) $ should be obviously positive
for small values of $\left| \mathbf{v-z}\right| $ and decrease as $\left| 
\mathbf{v-z}\right| $ increases, but as argued in section 3.1 we are here
concerned with the case when the $\mathbf{v}$\textbf{\ }integral over the
whole space is nil, that is 
\begin{equation}
\int C\left( \left| \mathbf{v-z}\right| \right) d^{3}v=\langle \psi \left| 
\hat{\rho}_{vac}\left( \mathbf{z}\right) \right| \psi \rangle =0,  \label{7a}
\end{equation}
see eq.$\left( \ref{3.5}\right) .$ Therefore $C\left( \left| \mathbf{v-z}%
\right| \right) $ will be negative for large values of $\left| \mathbf{v-z}%
\right| $. An illustrative example of a function with this behaviour is the
following 
\begin{eqnarray}
C\left( x\right) &=&an^{3}\exp \left( -3nx/b\right) -a\exp (-3x/b),n>>1
\label{8c} \\
&\rightarrow &C(0)=a(n^{3}-1),C\left( x\right) \simeq -a\exp (-3x/b)\text{
for }x>>b,  \nonumber
\end{eqnarray}
involving two parameters $\left\{ a,b\right\} .$ These parameters measure
roughly the size of the fluctuations and the the range of their correlation
function. The behaviour of the function C(x) suggest introducing an
auxiliary function $F(x)$ such that

\begin{equation}
C\left( x\right) =n^{3}F\left( nx\right) -F\left( x\right) ,  \label{8}
\end{equation}
where $n>>1$ is a real number and $F\left( x\right) $ is a positive function
of the argument that I assume rapidly decreasing at infinity, that is
fulfilling 
\begin{equation}
\lim_{x\rightarrow \infty }x^{3}F(x)=0\Rightarrow \lim_{x\rightarrow \infty
}x^{3}C(x)=0.  \label{8b}
\end{equation}
Eq.$\left( \ref{8}\right) $ guarantees that eq.$\left( \ref{7a}\right) $
holds true. Indeed for integrals over the whole space we have 
\begin{equation}
\int d^{3}xn^{3}F\left( nx\right) =\int d^{3}x^{\prime }F\left( x^{\prime
}\right) =\int d^{3}xF\left( x\right) \Rightarrow \int d^{3}xC\left(
x\right) =0.  \label{8a}
\end{equation}

Now we may evaluate eq.$\left( \ref{9}\right) $ taking eq.$\left( \ref{0.9}%
\right) $ into account. I start with the following $v$-integral of $C\left(
\left| \mathbf{v-z}\right| \right) $ 
\begin{equation}
I\equiv \int_{\left| \mathbf{v}\right| \mathbf{<}r}C\left( \left| \mathbf{v-z%
}\right| \right) d^{3}v=n^{3}\int_{\left| \mathbf{v}\right| \mathbf{<}%
r}F\left( n\left| \mathbf{v-z}\right| \right) d^{3}v-\int_{\left| \mathbf{v}%
\right| \mathbf{<}r}F\left( \left| \mathbf{v-z}\right| \right) d^{3}v.
\label{7b}
\end{equation}
In the limit $n\rightarrow \infty $ the function $n^{3}F\left( nx\right) $
becomes proportional to a 3D Dirac\'{}s delta $\delta ^{3}\left( x\right) $
as may be shown taking eq.$\left( \ref{8b}\right) $ into account. Thus for
very large $n$ the relevant contribution to the first integral of eq.$\left( 
\ref{7b}\right) $ comes from the region where $\left| \mathbf{v-z}\right| $
is small. Hence we may extend the $v$-integral to the whole space with fair
approximation provided that $\left| \mathbf{z}\right| <r,$ but neglect it if 
$\left| \mathbf{z}\right| >r.$ That is we may write 
\[
n^{3}\int_{\left| \mathbf{v}\right| \mathbf{<}r}F\left( n\left| \mathbf{v-z}%
\right| \right) d^{3}v\simeq \Theta \left( r-\left| \mathbf{z}\right|
\right) n^{3}\int_{\left| \mathbf{v}\right| \in \left( 0,\infty \right)
}F\left( n\left| \mathbf{v-z}\right| \right) d^{3}v, 
\]
where the step function $\Theta \left( y\right) =1$ if $y\geq 0$, $\Theta
\left( y\right) =0$ otherwise. Hence eq.$\left( \ref{7b}\right) $ gives 
\begin{eqnarray*}
I &\simeq &\Theta \left( r-\left| \mathbf{z}\right| \right)
n^{3}\int_{\left| \mathbf{v}\right| \in \left( 0,\infty \right) }F\left(
n\left| \mathbf{v-z}\right| \right) d^{3}v-\int_{\left| \mathbf{v}\right| 
\mathbf{<}r}F\left( \left| \mathbf{v-z}\right| \right) d^{3}v \\
&=&\Theta \left( r-z\right) n^{3}\int_{\left| \mathbf{x}\right| \in \left(
0,\infty \right) }F\left( nx\right) d^{3}x-\int_{\left| \mathbf{v}\right| 
\mathbf{<}r}F\left( \left| \mathbf{v-z}\right| \right) d^{3}v \\
&=&\Theta \left( r-z\right) \int_{\left| \mathbf{x}^{\prime }\right| \in
\left( 0,\infty \right) }F\left( x^{\prime }\right) d^{3}x^{\prime
}-\int_{\left| \mathbf{v}\right| \mathbf{<}r}F\left( \left| \mathbf{v-z}%
\right| \right) d^{3}v \\
&=&\Theta \left( r-z\right) \int_{\left| \mathbf{v}\right| \in \left(
0,\infty \right) }F\left( \left| \mathbf{v-z}\right| \right)
d^{3}v-\int_{\left| \mathbf{v}\right| \mathbf{<}r}F\left( \left| \mathbf{v-z}%
\right| \right) d^{3}v,
\end{eqnarray*}
leading to 
\begin{equation}
I=\Theta \left( r-z\right) \int_{\left| \mathbf{v}\right| \mathbf{\geq }%
r}F\left( \left| \mathbf{v-z}\right| \right) d^{3}v\text{ }-\Theta \left(
z-r\right) \int_{\left| \mathbf{v}\right| \mathbf{<}r}F\left( \left| \mathbf{%
v-z}\right| \right) d^{3}v\text{.}  \label{33}
\end{equation}
It is the case that I will integrate for $z\leq r$ everywhere in the rest of
this section whence eq.$\left( \ref{33}\right) $ becomes 
\[
I=\Theta \left( r-z\right) \int_{\left| \mathbf{v}\right| \mathbf{\geq }%
r}F\left( \left| \mathbf{v-z}\right| \right) d^{3}v\text{ } 
\]
in the following.

We get, taking eqs.$\left( \ref{9}\right) $ and $\left( \ref{0.9}\right) $
into acount, 
\begin{eqnarray}
J &\equiv &\int_{v\geq r,z<r}C\left( \left| \mathbf{v-z}\right| \right)
d^{3}vd^{3}z=4\int_{z<r}d^{3}z\int_{v>r}d^{3}vF\left( \left| \mathbf{v-z}%
\right| \right)  \nonumber \\
&=&32\pi ^{2}G^{2}r^{-2}\int_{0}^{r}z^{2}dz\int_{r}^{\infty
}v^{2}dv\int_{-1}^{1}duF\left( \sqrt{v^{2}+z^{2}-2vzu}\right) ,  \label{20}
\end{eqnarray}
where $u\equiv \cos \theta $, $\theta $ being the angle between the vectors $%
\mathbf{v}$ and $\mathbf{z}$. We know neither the two-point correlation
function $C\left( \left| \mathbf{v-z}\right| \right) $ nor $F\left( \left| 
\mathbf{v-z}\right| \right) $ in detail but I propose to characterize the
latter by just two parameters (see eq.$\left( \ref{8c}\right) )$, namely the
size $D$ and the range $\gamma .$ That is I will approximate the angular
integral in eq.$\left( \ref{20}\right) $ as follows 
\begin{equation}
f\equiv \int_{-1}^{1}duF\left( \sqrt{v^{2}+z^{2}-2vzu}\right) \approx
D\Theta \left( \gamma -\left| v-z\right| \right) ,  \label{19}
\end{equation}
where $\Theta \left( x\right) $ is the step function and we assume that the
parameter $\gamma >0$ is small in the sense that $\gamma <<r$ $.$ Thus we
get 
\begin{equation}
g_{rr}^{vac}\simeq 32\pi ^{2}G^{2}r^{-2}D\int_{r-\gamma
}^{r}z^{2}dz\int_{r}^{z+\gamma }v^{2}dv.  \label{21}
\end{equation}
For later convenience I will summarize the steps going from eq.$\left( \ref
{20}\right) $ to eq.$\left( \ref{21}\right) $, writing the following,
slightly more general, relation valid for any $\alpha \left( v,z\right) ,$%
\begin{equation}
\int_{v<r,z<r}C\left( \left| \mathbf{v-z}\right| \right) \alpha \left(
v,z\right) d^{3}vd^{3}z=8\pi ^{2}D\int_{r-\gamma
}^{r}z^{2}dz\int_{r}^{z+\gamma }\alpha \left( v,z\right) v^{2}dv.  \label{C}
\end{equation}

The integrals in eq.$\left( \ref{21}\right) $ are trivial and we obtain 
\begin{equation}
g_{rr}^{vac}\simeq \frac{G^{2}}{r^{2}}\times 32\pi ^{2}D\int_{r-\gamma
}^{r}z^{2}dz\left[ \frac{\left( z+\gamma \right) ^{3}}{3}-\frac{r^{3}}{3}%
\right] =16\pi ^{2}G^{2}D\gamma ^{2}r^{2}+O\left( \gamma ^{3}\right) .
\label{21a}
\end{equation}
The ratio $\gamma /r<<1$ is small because $\gamma $ is a length typical of
quantum fluctuations while $r$ is of order the typical distance amongs
galaxies (see comment after eq.$\left( \ref{30}\right) )$. Therefore we may
neglect terms of order $\gamma ^{3}$ whence we get 
\begin{equation}
g_{rr}^{vac}\simeq 16\pi ^{2}G^{2}Kr^{2},K\equiv D\gamma ^{2},  \label{22}
\end{equation}
where I have substituted the single parameter $K$ for the product $D$ times $%
\gamma ^{2}$. In the following I take the constant $K$ as the relevant
parameter, avoiding any detail about its origin from the two-point
correlation of vacuum fluctuations $C\left( \left| \mathbf{v-z}\right|
\right) .$

The terms of order $O\left( G^{2}\right) $ of $g_{tt}$, eq.$\left( \ref{13}%
\right) ,$ may be obtained in a way similar to those of $g_{rr}$. For the
first term we get 
\begin{eqnarray*}
c_{1} &\equiv &\left\langle \psi \left| \hat{c}_{1}\right| \psi
\right\rangle =4\int_{0}^{r}x^{-3}dx\left\langle \psi \left| \hat{m}\left(
x\right) ^{2}\right| \psi \right\rangle \\
&=&4\int_{0}^{r}x^{-3}dx\int_{0}^{x}d^{3}z\int_{0}^{x}d^{3}vC\left( \left| 
\mathbf{v-z}\right| \right) ,
\end{eqnarray*}
where $C\left( \left| \mathbf{v-z}\right| \right) $ is the correlation
function eq.$\left( \ref{8}\right) $. I will perform firstly the $x$
integral, that is 
\begin{eqnarray*}
c_{1} &=&4\int_{z<r}d^{3}z\int_{v<r}d^{3}vC\left( \left| \mathbf{v-z}\right|
\right) \int_{\max \left( v,z\right) }^{r}x^{-3}dx \\
&=&2\int_{z<r}d^{3}z\int_{v<r}d^{3}vC\left( \left| \mathbf{v-z}\right|
\right) (\frac{1}{\max \left( v,z\right) ^{2}}-\frac{1}{r^{2}}) \\
&=&16\pi ^{2}Dr^{-2}\int_{r-\gamma }^{r}z^{2}dz\left\{ \frac{1}{3}\left[
\left( z+\gamma \right) ^{3}-r^{3}\right] -r^{2}\left( z+\gamma -r\right)
\right\} ,
\end{eqnarray*}
where I have take eq.$\left( \ref{C}\right) $ into account. The result is
that $c_{1}$ is of order $O\left( \gamma ^{3}\right) $ whence this term
contributes but slightly to $g_{tt}.$

In order to get c$_{2}$ I start performing the $x$ and $y$ integrals, that
is 
\begin{eqnarray*}
c_{2} &\equiv
&8\int_{0}^{r}x^{-2}dx\int_{0}^{r}y^{-2}dy\int_{z<x}d^{3}z\int_{v<y}d^{3}vC%
\left( \left| \mathbf{v-z}\right| \right) \\
&=&\int_{z<r}d^{3}z\int_{v<r}d^{3}vC\left( \left| \mathbf{v-z}\right|
\right) \int_{z}^{r}x^{-2}dx\int_{v}^{r}y^{-2}dy \\
&=&\int_{z<r}\left( \frac{1}{z}-\frac{1}{r}\right) d^{3}z\int_{v<r}\left( 
\frac{1}{v}-\frac{1}{r}\right) d^{3}vC\left( \left| \mathbf{v-z}\right|
\right) .
\end{eqnarray*}
Taking eq.$\left( \ref{C}\right) $ into account we obtain 
\begin{eqnarray*}
c_{2} &=&8\pi ^{2}Dr^{-2}\int_{r-\gamma }^{r}z(r-z)\left[ \frac{1}{2}r\left(
\left( z+\gamma \right) ^{2}-r^{2}\right) -\frac{1}{3}\left( \left( z+\gamma
\right) ^{3}-r^{3}\right) \right] dz \\
&=&8\pi ^{2}Dr^{2}\gamma ^{2}+O\left( \gamma ^{3}\right) \simeq 8\pi
^{2}r^{2}K.
\end{eqnarray*}

Now we must compute the $G^{2}$ contribution to $g_{tt}$ coming from the
pressure operator $\hat{p}_{vac}\left( \mathbf{r,}t\right) $ of the vacuum
that is the terms $c_{3},c_{4}$ and $c_{5}.$ Before proceeding I must deal
with a difficulty due to the fact that eqs.$\left( \ref{4.0}\right) $ are
just valid for spherical symmetry. Actually that symmetry holds neither for
the distribution of matter in the region of interest nor for the
stress-energy of the quantum vacuum. In fact the stress-energy appears in
the form of localized operators of energy density $\hat{\rho}_{vac}\left( 
\mathbf{r,}t\right) $ and pressure $\hat{p}_{vac}\left( \mathbf{r,}t\right) $%
. Actually this was also the case of the mass and pressure distribution
leading the the terms of order $G$ in the metric elements in section 3.
Indeed I have solved the problem via a standard approximation that consists
of averaging the matter over the entire region. For the vacuum operator $%
\hat{\rho}_{vac}\left( \mathbf{r,}t\right) $ the problem is not too serious
because that operator enters just in the mass $\hat{m}(r),$ whose definition
in eq.$\left( \ref{7.0}\right) $ already involves an integral. However there
is a more difficult problem with the pressure operator $\hat{p}_{vac}$ that
actually depends on the position $\mathbf{x}$ rather than on the radial
coordinate $x$ alone as in eq.$\left( \ref{13}\right) .$ A plausible
approximation is to average the operator over the angular variables. Then I
will use $\hat{P}\left( x\right) ,$ an angular average operator, rather than 
$\hat{p}\left( x\right) ,$ in eq.$\left( \ref{13}\right) ,$ that is 
\begin{equation}
\hat{P}(x)\rightarrow \frac{1}{4\pi }\int_{0}^{\pi }\sin \theta d\theta
\int_{0}^{2\pi }d\phi \hat{p}\left( \mathbf{x}\right) =\frac{1}{4\pi x^{2}}%
\int \hat{p}\left( \mathbf{z}\right) d^{3}z\delta \left( x-z\right) ,
\label{P}
\end{equation}
$\delta \left( {}\right) $ being Dirac delta so that the $\mathbf{z}$
integral may be extended to the whole space with fair approximation.

After substituting $\hat{P}$ for $\hat{p}$ in eq.$\left( \ref{13}\right) $
we may get the expectation of the term of order $O(G^{2})$ belonging to the
metric element $\hat{g}_{tt}\left( r\right) .$ In order to compute the
numerical value we must introduce two new correlation functions similar to
eq.$\left( \ref{0.9}\right) $, that is 
\begin{eqnarray}
\frac{1}{2}\langle \psi \left| \hat{p}_{vac}\left( \mathbf{v}\right) \text{ }%
\hat{p}_{vac}\left( \mathbf{z}\right) +\text{ }\hat{p}_{vac}\left( \mathbf{z}%
\right) \hat{p}_{vac}\left( \mathbf{v}\right) \text{ }\right| \psi \rangle
&=&C_{pp}(\left| \mathbf{v-z}\right| ),  \nonumber \\
\frac{1}{2}\langle \psi \left| \hat{\rho}_{vac}\left( \mathbf{v}\right) 
\text{ }\hat{p}_{vac}\left( \mathbf{z}\right) +\text{ }\hat{p}_{vac}\left( 
\mathbf{z}\right) \hat{\rho}_{vac}\left( \mathbf{v}\right) \right| \psi
\rangle &=&C_{\rho p}(\left| \mathbf{v-z}\right| ).  \label{q}
\end{eqnarray}

The evaluation of the term c$_{3}$ is as follows, taking eq.$\left( \ref{13}%
\right) $, $\left( \ref{P}\right) $ and $\left( \ref{q}\right) $ into
account, 
\begin{eqnarray*}
c_{3} &=&32\pi ^{2}\int_{0}^{r}xdx\int_{0}^{r}ydy\langle \psi \left| \left[ 
\hat{p}\left( x\right) \hat{p}\left( y\right) +\hat{p}\left( y\right) \hat{p}%
\left( x\right) \right] \right| \psi \rangle \\
&\rightarrow &32\pi ^{2}\int_{0}^{r}xdx\int_{0}^{r}ydy\langle \psi \left| 
\hat{P}_{vac}\left( x\right) \text{ }\hat{P}_{vac}\left( y\right) +\text{ }%
\hat{P}_{vac}\left( y\right) \hat{P}_{vac}\left( x\right) \text{ }\right|
\psi \rangle \\
&=&64\pi ^{2}\int_{0}^{r}xdx\int_{0}^{r}ydy\frac{1}{16\pi ^{2}x^{2}y^{2}}%
\int d^{3}z\delta \left( x-z\right) \int d^{3}v\delta \left( y-v\right)
C_{pp}(\left| \mathbf{v-z}\right| ) \\
&=&4\int_{z<r}z^{-1}d^{3}z\int_{v<r}v^{-1}d^{3}vC_{pp}(\left| \mathbf{v-z}%
\right| ),
\end{eqnarray*}
where the $x$ and $y$ integrals have been performed.

Now I assume that an (approximate) equality holds similar to eq.$\left( \ref
{C}\right) $. Then I get 
\begin{eqnarray*}
c_{3} &=&32\pi ^{2}D_{pp}\int_{r-\gamma }^{r}zdz\int_{r}^{z+\gamma
}vdv=32\pi ^{2}D_{pp}\int_{r-\gamma }^{r}zdz\frac{\left( z+\gamma \right)
^{2}-r^{2}}{2} \\
&=&8\pi ^{2}\grave{D}_{pp}r^{2}\gamma ^{2}+O\left( \gamma ^{2}\right) \simeq
8\pi ^{2}r^{2}K_{pp},K_{pp}\equiv \grave{D}_{pp}\gamma ^{2}.
\end{eqnarray*}

Also I suppose that similar approximations are valid when the density
operator is combined with the pressure opertor. Thus we may calculate $c_{4}$
and $c_{5}$ in a similar way.

\begin{eqnarray*}
c_{4} &=&8\pi \int_{0}^{r}dx\langle \psi \left| \left[ \hat{m}\left(
x\right) \hat{p}\left( x\right) +\hat{p}\left( x\right) \hat{m}\left(
x\right) \right] \right| \psi \rangle \\
&\rightarrow &8\pi \int_{0}^{r}dx\langle \psi \left| \left[ \hat{m}\left(
x\right) \hat{P}\left( x\right) +\hat{P}\left( x\right) \hat{m}\left(
x\right) \right] \right| \psi \rangle \\
&=&16\pi \int_{0}^{r}dx\frac{1}{4\pi x^{2}}\int d^{3}z\delta \left(
x-z\right) \int_{v<x}d^{3}vC_{\rho p}(\left| \mathbf{v-z}\right| ) \\
&=&4\int_{z<r}z^{-2}d^{3}z\int_{v<r}d^{3}vC_{\rho p}(\left| \mathbf{v-z}%
\right| ) \\
&=&4D_{\rho p}\int_{r-\gamma }^{r}z^{-2}d^{3}z\int_{r}^{z+\gamma }d^{3}v \\
&=&64\pi ^{2}D_{\rho p}\int_{r-\gamma }^{r}dz\frac{\left( z+\gamma \right)
^{3}-r^{3}}{3}=32\pi ^{2}r^{2}D_{\rho p}\gamma ^{2}+O\left( \gamma
^{3}\right)
\end{eqnarray*}
Thus we get 
\[
c_{4}=32\pi ^{2}K_{\rho p}r^{2},K_{\rho p}\equiv D_{\rho p}. 
\]
\begin{eqnarray*}
c_{5} &=&8\pi \int_{0}^{r}x^{-2}dx\int_{0}^{r}ydy\langle \psi \left[ \hat{m}%
\left( x\right) \hat{p}\left( y\right) +\hat{p}\left( y\right) \hat{m}\left(
x\right) \right] \psi \rangle \\
&\rightarrow &16\pi \int_{0}^{r}x^{-2}dx\int_{0}^{r}ydy\int_{0}^{x}d^{3}z%
\frac{1}{4\pi y^{2}}\int d^{3}v\delta \left( y-v\right) C_{\rho p}(\left| 
\mathbf{v-z}\right| ) \\
&=&4\int_{0}^{r}\left( \frac{1}{z}-\frac{1}{r}\right)
d^{3}z\int_{v<r}v^{-1}d^{3}vC_{\rho p}(\left| \mathbf{v-z}\right| ) \\
&=&64\pi ^{2}r^{-1}D_{p\rho }\int_{r-\gamma }^{r}\left( r-z\right)
zdz\int_{r}^{z+\gamma }vdv \\
&=&32\pi ^{2}r^{-1}D_{p\rho }\int_{r-\gamma }^{r}\left( r-z\right) zdz\left[
\left( z+\gamma \right) ^{2}-r^{2}\right] =O\left( \gamma ^{3}\right) .
\end{eqnarray*}
Hence the term $c_{5}$ does not contribute to order $O\left( \gamma
^{2}\right) .$ In summary we have for the g$_{tt}$ element of the metric eq.$%
\left( \ref{30}\right) $

\begin{equation}
g_{tt}=1+\frac{4}{3}\pi G\rho _{mat}-\frac{8\pi G}{3}\rho _{DE}r^{2}+8\pi
^{2}G^{2}r^{2}(K+K_{pp}+4K_{\rho p}).  \label{7}
\end{equation}

It is plausible that the quantities $K$ and $K_{pp}$ are both positive but $%
K_{\rho P}$ negative. In fact we may assume that in quantum vacuum
fluctuations the pressure acts with a sign opposite to the mass density, in
agreement with the Lorentz invariant vacuum equation of state $p=-\rho .$
This suggests identifying 
\begin{equation}
K_{pp}=K,K_{\rho p}=-K  \label{7s}
\end{equation}
whence we get, taking eqs.$\left( \ref{9.0}\right) $ and $\left( \ref{22}%
\right) $, 
\begin{equation}
g_{rr}=1+\frac{8\pi G}{3}\left( \rho _{B}\left( t\right) +\rho _{DM}\left(
t\right) +\rho _{DE}\right) r^{2}+16\pi ^{2}G^{2}Kr^{2}.  \label{12}
\end{equation}
Similarly from eqs.$\left( \ref{7}\right) $ and $\left( \ref{7s}\right) $ we
obtain 
\begin{equation}
g_{tt}=1+\frac{8\pi G}{3}\left( \frac{1}{2}\rho _{B}\left( t\right) +\frac{1%
}{2}\rho _{DM}\left( t\right) -\rho _{DE}\right) r^{2}-16\pi ^{2}G^{2}Kr^{2}.
\label{12c}
\end{equation}
These results reproduce the standard ones eq.$\left( \ref{rofluct}\right) $
plus a correction due to the quantum vacuum fluctuations (i.e. the last term
in eqs.$\left( \ref{12}\right) $ and $\left( \ref{12c}\right) $).

\section{Results and discussion}

The main result of this article is that eqs.$\left( \ref{12}\right) $\ and $%
\left( \ref{12c}\right) $ should be substituted for the standard eqs.$\left( 
\ref{rofluct}\right) .$ Then the following should be substituted for eq.$%
\left( \ref{2.0}\right) $

\begin{eqnarray}
\rho _{DE}+6\pi GK &\simeq &(6.0\pm 0.2)\times 10^{-27}kg/m^{3}  \nonumber \\
&\Rightarrow &0<K\lesssim \frac{(6.0\pm 0.2)\times 10^{-27}kg/m^{3}}{6\pi G}.
\label{2.1}
\end{eqnarray}

The conclusion is that either the acceleration in the expansion of the
universe is due to the quantum vacuum (if the latter inequality is really an
equality) or the vacuum gives just a contribution to be added to the effect
of a dark energy. In the former case the value of the parameter K would be
following 
\begin{equation}
K\equiv D\gamma ^{2}=\frac{\rho _{DE}c^{2}}{6\pi G}\simeq 0.42kg^{2}/\text{m}%
^{4},\sqrt{K}\simeq 0.65kg/\text{m}^{2}.  \label{K}
\end{equation}
Taking eqs.$\left( \ref{7b}\right) $ to $\left( \ref{19}\right) $ into
account the quantity $\sqrt{K}$ may be seen as the product of the typical
mass density of the vacuum fluctuations $\sqrt{D}$ times its typical
correlation length $\gamma $. It is fitting that the value of $\sqrt{K},$ eq.%
$\left( \ref{K}\right) ,$ is not too far from the product of the typical
nuclear density, 2.3$\times $10$^{17}$ kg/m$^{3},$ times a typical nuclear
radius, about 10$^{-15}$m. For instance if the correlation length of the
vacuum energy density was 10$^{-11}$ m, a typical atomic distance, then the
fluctuation of the density would be about 10$^{-7}$ times the nuclear
density.

In summary our work does not prove that quantum vacuum fluctuations are a
valid alternative to dark energy. But it does show that such fluctuations
give rise to a contribution with qualitative effects similar to those of the
dark energy.

\end{document}